\documentclass[12pt]{article}

  \usepackage{graphicx}
  \usepackage{epsfig}
  \usepackage{amssymb}
  \usepackage{subfigure}

\evensidemargin - 1.truecm
\def\({\left(}
\def\){\right)}

\newcommand{\ep}{\varepsilon}
\newcommand{\Li}[2]{{\mbox{Li}}_{#1}\left(#2\right)}

\newcommand{\be}{\begin{equation}}
\newcommand{\ee}{\end{equation}}
\newcommand{\nn}{\nonumber}
\newcommand{\bea}{\begin{eqnarray}}
\newcommand{\eea}{\end{eqnarray}}
\newcommand{\bfig}{\begin{figure}}
\newcommand{\efig}{\end{figure}}
\newcommand{\bc}{\begin{center}}
\newcommand{\ec}{\end{center}}
\newcommand{\bd}{\begin{displaymath}}
\newcommand{\ed}{\end{displaymath}}

\newcommand{\sintheff}{\sin^2\theta_{\mbox{eff}}^{\mbox{lept}}}

\begin{document}


\begin{titlepage}
\nopagebreak
{\flushright{
        \begin{minipage}{4cm}
        Freiburg-THEP 05/02
        {\tt hep-ph/0504092}\\
        \end{minipage}        }

}
\vspace*{-1.5cm}                        
\vskip 3.5cm
\begin{center}
\boldmath
{\Large \bf Three-loop electroweak corrections to the $W$-boson mass
  and
  $\sintheff$
  in the large Higgs mass limit}
  \unboldmath
\vskip 1.2cm
{\large R.~Boughezal,\footnote{Email: 
{\tt Radja.Boughezal@physik.uni-freiburg.de}}
J.B.~Tausk\footnote{Email: 
{\tt Tausk@physik.uni-freiburg.de}}
and J.J.~van~der~Bij\footnote{Email: 
{\tt jochum@physik.uni-freiburg.de}}
} \\[2mm] 

\vskip .7cm
{\it  Fakult\"at f\"ur Mathematik und Physik, 
Albert-Ludwigs-Universit\"at
Freiburg, \\ D-79104 Freiburg, Germany} 
\vskip .3cm
 
\end{center}
\vskip 1.5cm


\begin{abstract} 

We present an analytical calculation of the leading three-loop
radiative correction to the $S$-parameter in the
Standard Model in the large Higgs mass limit.
Numerically,
$S^{(3)} =  1.1105 \times g^4/(1024\pi^3) \times m_H^4/M_W^4$.
When combined with the corresponding three-loop correction
to the $\rho$-parameter, this leads to shifts of
$\Delta^{(3)}\sintheff = 4.6\times 10^{-9}\times m_H^4/M_W^4$
in the effective weak mixing angle and
$\Delta^{(3)} M_W  = -6.3\times 10^{-4}\mbox{MeV}\times m_H^4/M_W^4$
in the $W$ boson mass. For both of these observables, the sign
of the three-loop correction is equal to that of the one-loop
correction.

\vskip .7cm 
\flushright{
        \begin{minipage}{12.3cm}
        \end{minipage}        }
\end{abstract}
\vfill
\end{titlepage}    


\section{Introduction}
\label{sec:intro}

Radiative corrections to four-fermion processes are sensitive to the
presence of heavy particles in the standard model. This allows one to
draw conclusions about the mass of the Higgs boson from fits of precise
electroweak measurements to theoretical predictions of the standard
model~\cite{LEP:2004}. For a light Higgs boson, the Higgs mass dependence
of the theoretical predictions is mainly due to one-loop radiative
corrections to the gauge boson propagators. These one-loop corrections
depend on $m_H$ logarithmically~\cite{Veltman:1976rt}. However,
because the Higgs self-interaction $\lambda$ is proportional to
$m_H^2$, there are higher order radiative corrections which grow like
powers of $m_H$ in the limit $m_H\to\infty$, eventually overcoming
their relative suppression by powers of $\alpha$. Such higher order
corrections could become important if the Higgs boson is very heavy.
At the two-loop level, the leading corrections are proportional to
$m_H^2$, but the numerical coefficient of these terms turns out to
be very small~\cite{vanderBij:1983bw,vanderBij2,Barbieri:1993ra},
and therefore, they are not important for $m_H$ less than a few TeV.
However, it has been suggested that the smallness of the two-loop corrections
may be somewhat accidental~\cite{akhoury}. If this is true, then one
may expect larger corrections to appear at the three loop level.

An explicit calculation of the leading three-loop correction
to the electroweak $\rho$-parameter in the large Higgs mass
limit~\cite{Boughezal:2004ef} has shown that, for this observable, the
numerical coefficient of the three-loop correction is indeed larger than the
two-loop coefficient, and that the leading three-loop correction
$\Delta\rho^{(3)}$ is already equal in magnitude to the two-loop
correction $\Delta\rho^{(2)}$ at a Higgs mass of approximately 480~GeV.
The purpose of the present paper is to extend our previous investigation
to other electroweak observables, in particular the sine of the effective
leptonic weak mixing angle
$\sintheff$,
defined in terms of the couplings of the $Z$-boson to leptons,
and the mass of the $W$ boson.
At the two-loop level, the complete electroweak fermionic corrections to
$\sintheff$ are known~\cite{Awramik:2004ge}, and for the $W$ mass,
both the fermionic and the bosonic corrections have been
calculated~\cite{Awramik:2003rn}. Here, we are concerned with the
leading three-loop bosonic corrections, which grow like
$m_H^4$ in the large Higgs mass limit.

We will present our results in terms of the $S$, $T$ and $U$ parameters
introduced by Peskin and Takeuchi~\cite{Peskin} to describe the effects
of heavy particles that enter only through corrections to the gauge
boson propagators (oblique corrections). They are defined in
terms of the transverse gauge boson self-energies at zero momentum
$\Sigma^{X}_T
 \equiv \Sigma^{X}_T(p^2)|_{p^2=0}$
and their first derivative
$\Sigma^{\prime X}_T
 \equiv \frac{\partial}{\partial p^2}\Sigma^{X}_T(p^2)|_{p^2=0}$
as
\begin{eqnarray}
\label{eq:Sdef}
S &\equiv& \frac{4 s_W^2 c_W^2}{\alpha}
  \left( \Sigma^{\prime ZZ}_T
       - \frac{c_W^2-s_W^2}{c_W s_W} \Sigma^{\prime AZ}_T
       - \Sigma^{\prime AA}_T \right)
\\
\label{eq:Tdef}
T &\equiv& \frac{1}{\alpha M_W^2}
  \left( c_W^2 \Sigma^{ZZ}_T
             - \Sigma^{WW}_T
  \right)
\\
\label{eq:Udef}
U &\equiv& \frac{4 s_W^2}{\alpha}
  \left( \Sigma^{\prime WW}_T
       - c_W^2 \Sigma^{\prime ZZ}_T
       - 2 c_W s_W  \Sigma^{\prime AZ}_T
       - s_W^2 \Sigma^{\prime AA}_T \right)
\end{eqnarray}
where $c_W=\cos\theta_W$, $s_W=\sin\theta_W$.
The leading corrections to
four-fermion processes in the heavy Higgs mass limit can be treated in
this framework provided that there are no further contributions from
vertex and box diagrams.
In the renormalization scheme we are using, where  $m_H$ and $M_W$
are defined by subtractions at $p^2=-m_H^2$ and $p^2=0$, respectively
(which is equivalent to the on-shell scheme if one is only interested
in the leading term in the large Higgs mass limit), this condition is
satisfied~\cite{vanderBij:1983bw,Einhorn:1988tc}.


\section{Calculation}
\label{sec:calculation}

All our diagrams are generated by the program QGRAF~\cite{qgraf}.
The rest of the calculation is done mainly in FORM~\cite{FORM}.
First of all, we need the bare gauge boson self-energies,
which we decompose into their transverse and longitudinal components
as
\begin{equation}
\Sigma^{X}_{\mu\nu}(p)
 = \left(g_{\mu\nu}-\frac{p_{\mu}p_{\nu}}{p^2}\right)
   \Sigma^{X}_T(p^2)
 + \frac{p_{\mu}p_{\nu}}{p^2} \,
   \Sigma^{X}_L(p^2) \, ,
\end{equation}
where $X = AA, AZ, ZZ, WW$.
The scalar functions $\Sigma^{X}_T(p^2)$ and $\Sigma^{X}_L(p^2)$
are extracted from $\Sigma^{X}_{\mu\nu}(p)$ by means of the projectors
$
P_T = \frac{1}{d-1} \left(g_{\mu\nu}-p_{\mu}p_{\nu}/p^2\right)
$ and
$
P_L = p_{\mu}p_{\nu}/p^2
$
respectively, where $d=4-\ep$ is the space-time dimension, and expanded in
a Taylor series up to order $p^2$. The coefficients are vacuum integrals,
which, in general, depend on three different non-zero masses: $m_H$, $M_W$
and $M_Z$. We use the method of expansion by regions~\cite{smirnovbook}
to expand them in powers of $m_H$ for $m_H\gg M_W, M_Z$, keeping all terms
proportional to $m_H^4$ or higher in the three-loop gauge boson self-energies.
As discussed in more detail in ref.~\cite{Boughezal:2004ef}, the expansion
by regions produces:\\
(a) three-loop single-scale vacuum integrals, where
the mass scale is $m_H$ (from the region where all loop momenta are large);\\
(b) products of one- (two-) loop vacuum integrals depending on $m_H$
and two- (one-) loop vacuum integrals depending on the small scales
$M_W$ and $M_Z$ (from the regions where some of the loop momenta are
small and others are large);\\
(c) and finally three-loop vacuum integrals depending on $M_W$ and $M_Z$
(from the region where all loop momenta are small).

Using integration-by-parts identities, we reduce all of these vacuum
integrals to master integrals. For cases (a) and (c), we use the automatic
integral reduction package AIR~\cite{Anastasiou:2004vj}. The integrals
that occur in case (a) have ${\cal N}_{+} \leq 5$ and ${\cal N}_{-} \leq
2$, where ${\cal N}_{+}$ is equal to the sum of the positive indices,
minus the number of positive indices, and ${\cal N}_{-}$ denotes the
absolute value of the sum of the negative indices.  The three-loop
single-scale master integrals appearing in case (a) can be found as
expansions in $\ep$ in ref.~\cite{Broadhurst:1998rz,Fleischer:1999mp}.

In case (c), we have to reduce two-scale integrals with ${\cal N}_{+}
\leq 4$ and ${\cal N}_{-} \leq 2$. However, once we sum over all diagrams,
in all the three-loop gauge boson self-energies, the non-factorizable
three-loop master integrals occurring in case (c) cancel, so that explicit
formulae for such master integrals are not needed.

The longitudinal parts of the gauge boson self-energies are related to
the self-energies of the scalars and the mixings between the scalars and
the gauge bosons by a set of Ward identities, which are shown explicitly
in appendix A. We have verified that these Ward identities are satisfied
by the full (ie., including tadpole contributions), unrenormalized
self-energies (up to order $p^2$).

\section{Renormalization}
\label{sec:Renormalization}
We work with the Lagrangian,
${\cal L}={\cal L}_{inv}+{\cal L}_{fix}+{\cal L}_{FP}$,
with the invariant part
\begin{equation}
\label{eq:Linv}
{\cal L}_{inv} = -\frac{1}{4} W^a_{\mu\nu} W^{a,\mu\nu}
           -\frac{1}{4} B_{\mu\nu} B^{\mu\nu}
           -\left( D_{\mu} \Phi \right)^{\dagger}
            \left( D^{\mu} \Phi \right)
           -\frac{1}{2} \lambda {\left(\Phi^{\dagger}\Phi\right)}^2 
           -\mu \, \Phi^{\dagger}\Phi \, ,
\end{equation}
with the Higgs doublet
\begin{equation}
\Phi= \frac{1}{\sqrt{2}} \left(
 \begin{array}{c}
 H+\sqrt{2}v + i \phi^0
\\
 i \phi^1 - \phi^2
 \end{array}
\right) \, ,
\end{equation}
and the gauge fixing term
\begin{equation}
\label{eq:Lfix}
{\cal L}_{fix} = -C^+ C^- - \frac{1}{2} {(C^Z)}^2 - \frac{1}{2} {(C^A)}^2 \, ,
\end{equation}
with
\begin{eqnarray}
\label{eq:Cpm}
C^{\pm} & = & - \partial_{\mu} W^{\pm,\mu} + M_W \phi^{\pm}
\\
\label{eq:Cz}
C^{Z} & = & - \partial_{\mu} Z^{\mu} + M_Z \phi^{0}
\\
C^{A} & = & - \partial_{\mu} A^{\mu} \, .
\end{eqnarray}
The parameters $v$, $\lambda$ and $\mu$ are given by:
\begin{eqnarray}
v & = & \sqrt{2} \frac{M_W}{g} \, ,
\\
\label{eq:lambdadef}
\lambda & = & g^2 \frac{m_H^2}{4 M_W^2}\, ,
\\
\label{eq:mudef}
\mu & = & \beta-\frac{1}{2} m_H^2\, .
\end{eqnarray}

Giving $\beta$ a non-zero value produces a term in the Lagrangian that is
linear in the Higgs field $H$. We adjust this parameter
order by order to make the renormalized Higgs tadpole
vanish. We introduce two further renormalization constants,
$Z_H$ and $Z_{m_H}$,
by making the following substitutions in 
${\cal L}_{inv}$ (but not in ${\cal L}_{fix}$ or ${\cal L}_{FP}$):
\begin{eqnarray}
\label{eq:mren}
m_H & \to & Z_{m_H} \, m_H
\\
\label{eq:Mren}
M_W & \to & Z_H \, M_W
\\
\label{eq:Hren}
H & \to & Z_H \, H
\\
\label{eq:phiren}
\phi^{\pm} & \to & Z_H \phi^{\pm}
\\
\label{eq:phi0ren}
\phi^{0} & \to & Z_H \phi^{0} \, .
\end{eqnarray}
The renormalization constants
are fixed by imposing the following conditions on the
renormalized Higgs tadpole $\Gamma^{H,ren}$ and on the
renormalized $\phi$ and $H$ self-energies:
\begin{eqnarray}
\Gamma^{(1),H,ren} & = & {\cal O}(m_H^0)
\\
\Gamma^{(2),H,ren} & = & {\cal O}(m_H^2)
\\
\Sigma^{\prime(1),\phi\phi,ren}|_{p^2=0} & = & {\cal O}(m_H^0)
\\
\Sigma^{\prime(2),\phi\phi,ren}|_{p^2=0} & = & {\cal O}(m_H^2)
\\
\mbox{Re} \, \Sigma^{(1),HH,ren}|_{p^2+m_H^2=0} & = & {\cal O}(m_H^2)
\\
\mbox{Re} \, \Sigma^{(2),HH,ren}|_{p^2+m_H^2=0} & = & {\cal O}(m_H^4)
\end{eqnarray}
These renormalizations remove all the terms of order $m_H^2$ and
of order $m_H^4$ from the one- and two-loop gauge boson self-energies,
the $\phi$ self-energies, and the mixings between $\phi$'s and
gauge bosons.
This ensures that no two- or three-loop vertex or box graphs containing
such self-energies as subgraphs can give corrections that grow
like $m_H^2$ or $m_H^4$ in the large Higgs mass limit, and
allows us to restrict our attention to the corrections that
come from the gauge boson self-energies.
The renormalization procedure used here is slightly different
from the one used in ref.~\cite{Boughezal:2004ef}, where the
parameter $\beta$ was not used and therefore, tadpole contributions
had to be taken into account explicitly. In particular, the
constant $Z_{m_H}$ used here is different from $Z_m$ of
ref.~\cite{Boughezal:2004ef}. However, the two procedures
are equivalent to each other.
Explicit formulae for the renormalization constants are listed
in appendix B.

We have verified that the renormalized longitudinal photon self-energy
and photon-Z mixing are zero, as they should.


\section{Results and conclusion}
\label{sec:result}

Here, we collect the leading one-loop~\cite{Antonelli,Veltman:1980fk}
and two-loop~\cite{vanderBij2,Barbieri:1993ra} contributions
and the new three-loop contribution to the $S$-parameter
in the heavy Higgs limit:
\begin{eqnarray}
\label{eq:Sres}
S^{(1)}  & = & \frac{1}{12\pi} 
               \log\left(\frac{m_H^2}{M_W^2}\right) \, ,
\\
S^{(2)}  & = & \frac{1}{4\pi} \left(\frac{g^2}{16\pi^2}\right)
                   \frac{m_H^2}{M_W^2}
 \left(\,
  - \frac{35}{72}
  - \frac{1}{8} \, \pi \sqrt{3}
  + \frac{7}{54} \, \pi^2
 \, \right) 
\nn\\
         & = & \frac{1}{4\pi} \left(\frac{g^2}{16\pi^2}\right)
                   \frac{m_H^2}{M_W^2}
                   \left(\, 0.1131 \, \right)  \,,
\\
S^{(3)}  & = & \frac{1}{4\pi} {\left(\frac{g^2}{16\pi^2}\right)}^2
                   \frac{m_H^4}{M_W^4}
 \left(\,
      \frac{1153}{576}
    - \frac{19}{48} \, \pi \sqrt{3}
    + \frac{13}{16} \, \pi C
    + \frac{2753}{10368} \, \pi^2
    - \frac{109}{432} \, \pi^3 \sqrt{3}
\right.\nn \\ && \hspace{2cm}
    - \frac{7199}{155520} \, \pi^4
    + \frac{7}{4} \, \sqrt{3} \, C \log{3}
    - \frac{21}{8} \, \sqrt{3} \, {\mbox{Ls}}_3(\frac{2 \pi}{3})
\nn \\ && \hspace{2cm} \left.
    - \frac{105}{16} \, \sqrt{3} \, C
    + \frac{38525}{3456} \, \zeta{(3)}
    - \frac{25}{24} \, C^2
    - \frac{17}{18} \, U_{3,1}
    - 2 \, V_{3,1}
 \, \right) 
\nn\\
         & = & \frac{1}{4\pi} {\left(\frac{g^2}{16\pi^2}\right)}^2
                   \frac{m_H^4}{M_W^4}
                   \left(\, 1.1105 \, \right) \,.
\end{eqnarray}
The constants appearing in these expressions
are~\cite{Broadhurst:1998rz,Fleischer:1999mp}:
\begin{eqnarray}
U_{3,1} &=& \frac{1}{2} \zeta(4) + \frac{1}{2} \zeta(2) \log^2 2
           - \frac{1}{12} \log^4 2 - \Li{4}{\frac{1}{2}} 
    \;\; = \;\; -0.11787599965\,, \hspace*{4mm}
\\
V_{3,1} &=& 
\sum_{m>n>0}\frac{(-1)^m\cos(2\pi n/3)}{m^3n}
    \;\; = \;\; -0.03901272636\,,
\\ C &=& {\mbox{Cl}}_2\left(\pi/3\right)
    \;\; = \;\; 1.0149416064\,,
\\{\mbox{Ls}}_3(\frac{2 \pi}{3}) &=&
- \int_0^{2 \pi/3} \mbox{d} \phi \;
\log^{2} \left| 2 \sin\frac{\phi}{2} \right|
    \;\; = \;\; -2.1447672126\,.
\end{eqnarray}
The $T$-parameter is related to the $\rho$-parameter by
$\Delta\rho = \alpha T$, so that, from
refs.~\cite{Longhitano:1980iz,vanderBij:1983bw,Boughezal:2004ef},
\begin{eqnarray}
\label{eq:Tres}
T^{(1)} & = & - \frac{3}{16\pi c_W^2} 
               \log\left(\frac{m_H^2}{M_W^2}\right) \, ,
\\
T^{(2)} & = & \frac{1}{4\pi c_W^2} \left(\frac{g^2}{16\pi^2}\right)
                   \frac{m_H^2}{M_W^2}
                   \left(\, 0.1499 \, \right)  \,,
\\
T^{(3)} & = & \frac{1}{4\pi c_W^2} {\left(\frac{g^2}{16\pi^2}\right)}^2
                   \frac{m_H^4}{M_W^4}
                   \left(\, -1.7282 \, \right) \,.
\end{eqnarray}
In contrast to $S$, the parameter $U$ can only be
different from zero if custodial symmetry is
broken, and as a result, it should be suppressed compared to
$S$~\cite{Barbieri:1993ra}. In the approximation
where we keep only logarithmic terms in $m_H$ at the one-loop
level, quadratic terms at the two-loop level and quartic
terms at the three-loop level, we find that $U$ vanishes.
This provides a very useful check on the calculation.

Knowing $S$, $T$, and $U$, one can easily obtain the
corresponding corrections to various electroweak
precision observables~\cite{Peskin,Marciano:1990dp,Altarelli:1990zd}.
We illustrate this by two examples.
The effective weak mixing angle is shifted relative
to its tree level value, expressed in terms of
$\alpha$, $G_F$ and $M_Z$, by
\begin{equation}
\sintheff =
\Delta\sintheff +
\frac{1}{2}
 -\sqrt{\frac{1}{4} - \frac{\pi\alpha}{\sqrt{2}\,G_F M_Z^2}}\, ,
\end{equation}
with
\begin{equation}
\label{eq:deltasin}
\Delta\sintheff =
\frac{\alpha}{c_W^2-s_W^2}
\left( \frac{1}{4} S - s_W^2 c_W^2 T \right) \, .
\end{equation}
Similarly, one finds a shift in the $W$-mass,
\begin{equation}
M_W = \Delta M_W +
 M_Z \sqrt{\frac{1}{2}
 +\sqrt{\frac{1}{4} - \frac{\pi\alpha}{\sqrt{2}\,G_F M_Z^2}}} \, ,
\end{equation}
with
\begin{equation}
\label{eq:deltamw}
\Delta M_W =
\frac{\alpha M_W}{2(c_W^2-s_W^2)}
\left( - \frac{1}{2} S + c_W^2 T + \frac{c_W^2-s_W^2}{4s_W^2} U \right) \, .
\end{equation}

\begin{table}
\[\begin{array}{|c|c|c|c|}  \hline  \rule{0pt}{14pt}
m_H/M_W
 & \Delta^{(1)}\sintheff
 & \Delta^{(2)}\sintheff
 & \Delta^{(3)}\sintheff
\\[2pt] \hline \rule{0pt}{14pt}
2  & 3.8\times 10^{-4} & -6.7\times 10^{-8} & 7.4\times 10^{-8}
\\
5  & 8.9\times 10^{-4} & -4.2\times 10^{-7} & 2.9\times 10^{-6}
\\
10 & 1.3\times 10^{-3} & -1.7\times 10^{-6} & 4.6\times 10^{-5}
\\
15 & 1.5\times 10^{-3} & -3.8\times 10^{-6} & 2.3\times 10^{-4}
\\
20 & 1.6\times 10^{-3} & -6.7\times 10^{-6} & 7.4\times 10^{-4}
\\
25 & 1.8\times 10^{-3} & -1.1\times 10^{-5} & 1.8\times 10^{-3}
\\ \hline
\end{array}\]
\caption{Corrections to $\sintheff$ as a function 
         of \,\,${{m_H}/{M_W}}$.}
\label{tab:sinth}
\end{table}

\begin{table}
\[\begin{array}{|c|c|c|c|}  \hline  \rule{0pt}{14pt}
m_H/M_W
 & \Delta^{(1)}M_W
 & \Delta^{(2)}M_W
 & \Delta^{(3)}M_W
\\[2pt] \hline \rule{0pt}{14pt}
2  & -0.055 &  0.000041 & -0.000010
\\
5  & -0.13  &  0.00025  & -0.00039
\\
10 & -0.18  &  0.0010   & -0.0063
\\
15 & -0.21  &  0.0023   & -0.032
\\
20 & -0.24  &  0.0041   & -0.10
\\
25 & -0.26  &  0.0064   & -0.25
\\ \hline
\end{array}\]
\caption{Corrections to $M_W$ in GeV as a function of \,\,${{m_H}/{M_W}}$.}
\label{tab:MW}
\end{table}

Numerical values of these shifts, calculated using
\begin{equation}
g^2 = \frac{e^2}{s_W^2} = \frac{4\pi\alpha}{s_W^2}
\end{equation}
for the weak coupling constant, with
$\alpha=1/137$
and
$s_W^2 = 0.23$,
are shown in Tables~\ref{tab:sinth} and \ref{tab:MW}
and in Figures~{\ref{fig:sinth} and \ref{fig:mwshift}.
The two-loop corrections are extremely small. This is partly due to a
cancellation between $S^{(2)}$ and $T^{(2)}$ in eqs.~(\ref{eq:deltasin})
and (\ref{eq:deltamw}). At the three-loop level, there is no such
cancellation between $S^{(3)}$ and $T^{(3)}$, since they have opposite
signs.
For both observables, the three-loop correction becomes
equal to the one-loop correction at $m_H\approx 2~\mbox{TeV}$.
Since the three-loop term has the same sign as the
one-loop term in both cases, it is clearly not
possible for the radiative corrections due to a heavy
Higgs boson to mimic those of a light Higgs boson.

\begin{figure}
\begin{center}
\begin{picture}(0,0)%
\includegraphics{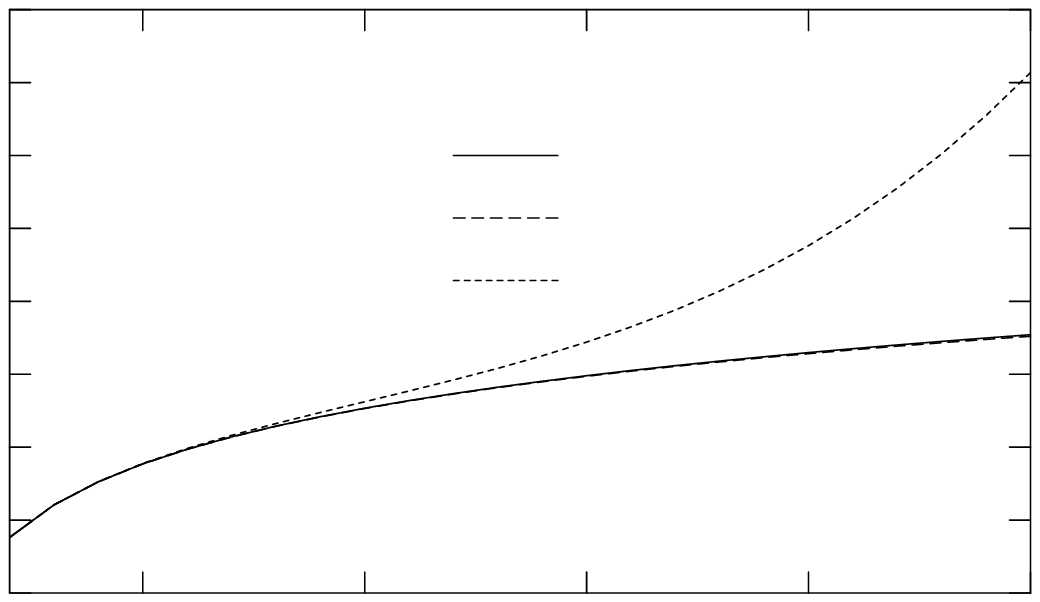}%
\end{picture}%
\setlength{\unitlength}{0.0200bp}%
\begin{picture}(18000,10800)(0,0)%
\put(2100,1800){\makebox(0,0)[r]{\strut{}0.0}}%
\put(2100,2850){\makebox(0,0)[r]{\strut{}0.5}}%
\put(2100,3900){\makebox(0,0)[r]{\strut{}1.0}}%
\put(2100,4950){\makebox(0,0)[r]{\strut{}1.5}}%
\put(2100,6000){\makebox(0,0)[r]{\strut{}2.0}}%
\put(2100,7050){\makebox(0,0)[r]{\strut{}2.5}}%
\put(2100,8100){\makebox(0,0)[r]{\strut{}3.0}}%
\put(2100,9150){\makebox(0,0)[r]{\strut{}3.5}}%
\put(2100,10200){\makebox(0,0)[r]{\strut{}4.0}}%
\put(4317,1200){\makebox(0,0){\strut{} 5}}%
\put(7513,1200){\makebox(0,0){\strut{} 10}}%
\put(10709,1200){\makebox(0,0){\strut{} 15}}%
\put(13904,1200){\makebox(0,0){\strut{} 20}}%
\put(17100,1200){\makebox(0,0){\strut{} 25}}%
\put(600,6000){\rotatebox{90}{\makebox(0,0){\strut{}$\Delta \sintheff\, (10^{-3})$}}}%
\put(9750,300){\makebox(0,0){\strut{}$m_H/M_W$}}%
\put(8491,8100){\makebox(0,0)[r]{\strut{}$\Delta^{(1)}$}}%
\put(8491,7200){\makebox(0,0)[r]{\strut{}$\Delta^{(1)}+\Delta^{(2)}$}}%
\put(8491,6300){\makebox(0,0)[r]{\strut{}$\Delta^{(1)}+\Delta^{(2)}+\Delta^{(3)}$}}%
\end{picture}%
\end{center}
\caption{
Shifts in $\sintheff$ as a function of $m_H/M_W$.}
\label{fig:sinth}
\end{figure}

\begin{figure}
\begin{center}
\begin{picture}(0,0)%
\includegraphics{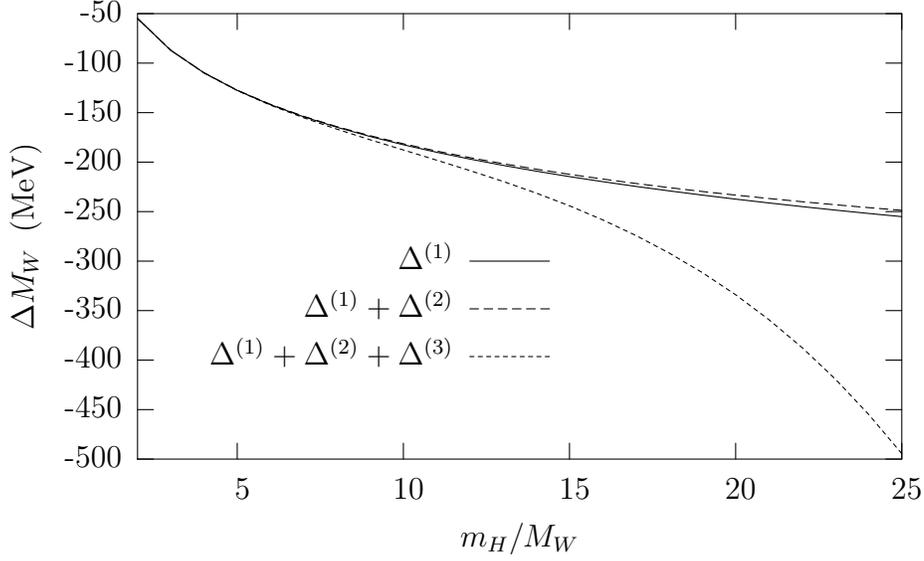}%
\end{picture}%
\setlength{\unitlength}{0.0200bp}%
\begin{picture}(18000,10800)(0,0)%
\put(2400,1800){\makebox(0,0)[r]{\strut{}-500}}%
\put(2400,2733){\makebox(0,0)[r]{\strut{}-450}}%
\put(2400,3667){\makebox(0,0)[r]{\strut{}-400}}%
\put(2400,4600){\makebox(0,0)[r]{\strut{}-350}}%
\put(2400,5533){\makebox(0,0)[r]{\strut{}-300}}%
\put(2400,6467){\makebox(0,0)[r]{\strut{}-250}}%
\put(2400,7400){\makebox(0,0)[r]{\strut{}-200}}%
\put(2400,8333){\makebox(0,0)[r]{\strut{}-150}}%
\put(2400,9267){\makebox(0,0)[r]{\strut{}-100}}%
\put(2400,10200){\makebox(0,0)[r]{\strut{}-50}}%
\put(4578,1200){\makebox(0,0){\strut{} 5}}%
\put(7709,1200){\makebox(0,0){\strut{} 10}}%
\put(10839,1200){\makebox(0,0){\strut{} 15}}%
\put(13970,1200){\makebox(0,0){\strut{} 20}}%
\put(17100,1200){\makebox(0,0){\strut{} 25}}%
\put(600,6000){\rotatebox{90}{\makebox(0,0){\strut{}$\Delta M_W\,$ (MeV)}}}%
\put(9900,300){\makebox(0,0){\strut{}$m_H/M_W$}}%
\put(8661,5533){\makebox(0,0)[r]{\strut{}$\Delta^{(1)}$}}%
\put(8661,4633){\makebox(0,0)[r]{\strut{}$\Delta^{(1)}+\Delta^{(2)}$}}%
\put(8661,3733){\makebox(0,0)[r]{\strut{}$\Delta^{(1)}+\Delta^{(2)}+\Delta^{(3)}$}}%
\end{picture}%
\end{center}
\caption{
Shifts in $M_W$ as a function of $m_H/M_W$.}
\label{fig:mwshift}
\end{figure}


\subsection*{Acknowledgements}
This work was supported by the DFG within the project
"(Nicht)-perturbative Quantenfeldtheorie".

\appendix


\section{Ward identities}

The Ward identities for the self-energies can be derived by expanding
the following identities for the dressed propagators $G$:
\begin{eqnarray}
\label{eq:WIGaa}
1 & = & p^{\mu} p^{\nu} G^{AA}_{\mu\nu}(p)
\\
\label{eq:WIGaz}
0 & = & p^{\mu} p^{\nu} G^{AZ}_{\mu\nu}(p)
 + i p^{\mu} M_Z G^{\phi^0 A}_{\mu}(p)
\\
\label{eq:WIGzz}
1 & = & p^{\mu} p^{\nu} G^{ZZ}_{\mu\nu}(p)
 + 2 i p^{\mu} M_Z G^{\phi^0 Z}_{\mu}(p)
 + M_Z^2 G^{\phi^0\phi^0}(p)
\\
\label{eq:WIGww}
1 & = & p^{\mu} p^{\nu} G^{WW}_{\mu\nu}(p)
 + 2 i p^{\mu} M_W G^{\phi W}_{\mu}(p)
 + M_W^2 G^{\phi\phi}(p)
\end{eqnarray}
Expanding the propagators $G$ in eqs.~(\ref{eq:WIGaa})--(\ref{eq:WIGww})
in terms of the self-energies $\Sigma$, one obtains a set of
Ward identities.
For the mixings between Goldstone and gauge bosons,
we write
\begin{equation}
 \Sigma^{X}_{\mu}(p) = p_{\mu} \Sigma^{X}(p^2)
\end{equation}
where $X= \phi^0 A, \phi^0 Z, \phi W$.

Up to three loop order, we find the following relations.
\begin{eqnarray}
0 & = & \Sigma^{(1),AA}_L
\\
0 & = &\Sigma^{(1),AZ}_L
+ i M_Z \Sigma^{(1),\phi^0 A}
\\
0 & = & p^2 \Sigma^{(1),ZZ}_L
+ 2 i M_Z p^2 \Sigma^{(1),\phi^0 Z}
+ M_Z^2 \Sigma^{(1),\phi^0\phi^0}
\\
0 & = & p^2 \Sigma^{(1),WW}_L
+ 2 i M_W p^2 \Sigma^{(1),\phi W}
+ M_W^2 \Sigma^{(1),\phi\phi}
\\
0 & = &
         \Sigma^{(2),A A}_L
          - {\left(\Sigma^{(1),\phi^0 A}\right)}^2
\\
0 & = &
         \Sigma^{(2),A Z}_L
          - \Sigma^{(1),\phi^0 A} \Sigma^{(1),\phi^0 Z}
       + i M_Z  \left(
            \frac{1}{p^2} \Sigma^{(1),\phi^0 \phi^0} \Sigma^{(1),\phi^0 A}
          + \Sigma^{(2),\phi^0 A}
         \right)
\\
0 & = &
         p^2 \Sigma^{(2),Z Z}_L 
          - p^2 {\left(\Sigma^{(1),\phi^0 Z}\right)}^2
       + 2\, i M_Z  \left(
            \Sigma^{(1),\phi^0 \phi^0} \Sigma^{(1),\phi^0 Z}
          +  p^2 \Sigma^{(2),\phi^0 Z} 
         \right)
\nn \\ & + &
         M_Z^2  \left(
            \frac{1}{p^2} {\left(\Sigma^{(1),\phi^0 \phi^0}\right)}^2
          + \Sigma^{(2),\phi^0 \phi^0}
         \right)
\\
0 & = &
         p^2 \Sigma^{(2),W W}_L 
          - p^2 {\left(\Sigma^{(1),\phi W}\right)}^2
       + 2\, i M_W  \left(
            \Sigma^{(1),\phi \phi} \Sigma^{(1),\phi W}
          +  p^2 \Sigma^{(2),\phi W} 
         \right)
\nn \\ & + &
         M_W^2  \left(
            \frac{1}{p^2} {\left(\Sigma^{(1),\phi \phi}\right)}^2
          + \Sigma^{(2),\phi \phi}
         \right)
\\
0 & = &
            \Sigma^{(3),A A}_L
       - \frac{1}{p^2} \Sigma^{(1),\phi^0 \phi^0}
                       {\left(\Sigma^{(1),\phi^0 A}\right)}^2
          - 2\, \Sigma^{(1),\phi^0 A} \Sigma^{(2),\phi^0 A}
\\
0 & = &
            \Sigma^{(3),A Z}_L
       -  \frac{1}{p^2} \Sigma^{(1),\phi^0 \phi^0}
                         \Sigma^{(1),\phi^0 A} \Sigma^{(1),\phi^0 Z}
          - \Sigma^{(1),\phi^0 A} \Sigma^{(2),\phi^0 Z}
          - \Sigma^{(1),\phi^0 Z} \Sigma^{(2),\phi^0 A}
\nn \\ & + &
         i M_Z  \left(
            \frac{1}{(p^2)^2} {\left(\Sigma^{(1),\phi^0 \phi^0}\right)}^2
                               \Sigma^{(1),\phi^0 A}
          +  \frac{1}{p^2} \Sigma^{(1),\phi^0 \phi^0} \Sigma^{(2),\phi^0 A}
          +  \frac{1}{p^2} \Sigma^{(2),\phi^0 \phi^0} \Sigma^{(1),\phi^0 A}
\right. \nn \\ && \left.
          {} + \Sigma^{(3),\phi^0 A}
         \right)
\\
0 & = & p^2 \Sigma^{(3),Z Z}_L
          - \Sigma^{(1),\phi^0 \phi^0} {\left(\Sigma^{(1),\phi^0 Z}\right)}^2
          - 2\, p^2 \Sigma^{(1),\phi^0 Z} \Sigma^{(2),\phi^0 Z}
\nn \\
      & + & 2\, i M_Z  \left(
          \frac{1}{p^2} {\left(\Sigma^{(1),\phi^0 \phi^0}\right)}^2
            \Sigma^{(1),\phi^0 Z}
          + \Sigma^{(1),\phi^0 \phi^0} \Sigma^{(2),\phi^0 Z}
          + \Sigma^{(2),\phi^0 \phi^0} \Sigma^{(1),\phi^0 Z}
\right. \nn \\ && \left.
          {} + p^2 \Sigma^{(3),\phi^0 Z}
          \right)
\nn \\
      & + & M_Z^2  \left(
            \frac{1}{(p^2)^2} {\left(\Sigma^{(1),\phi^0 \phi^0}\right)}^3
          + \frac{2}{p^2} \Sigma^{(1),\phi^0 \phi^0} \Sigma^{(2),\phi^0 \phi^0}
          + \Sigma^{(3),\phi^0 \phi^0}
          \right)
\\
0 & = & p^2 \Sigma^{(3),W W}_L
          - \Sigma^{(1),\phi \phi} {\left(\Sigma^{(1),\phi W}\right)}^2
          - 2\, p^2 \Sigma^{(1),\phi W} \Sigma^{(2),\phi W}
\nn \\
      & + & 2\, i M_W  \left(
          \frac{1}{p^2} {\left(\Sigma^{(1),\phi \phi}\right)}^2
            \Sigma^{(1),\phi W}
          + \Sigma^{(1),\phi \phi} \Sigma^{(2),\phi W}
          + \Sigma^{(2),\phi \phi} \Sigma^{(1),\phi W}
\right. \nn \\ && \left.
         {} + p^2 \Sigma^{(3),\phi W}
          \right)
\nn \\ & + &
            M_W^2  \left(
            \frac{1}{(p^2)^2} {\left(\Sigma^{(1),\phi \phi}\right)}^3
          + \frac{2}{p^2} \Sigma^{(1),\phi \phi} \Sigma^{(2),\phi \phi}
          + \Sigma^{(3),\phi \phi}
          \right)
\end{eqnarray}


\section{Renormalization constants}
In $d=4-\ep$ dimensions, the renormalization constants
$Z_H$ and $\beta$ are given by
$Z_H=1-\delta_H^{(1)}-\delta_H^{(2)}$
and
$\beta = m_H^2 \left( \beta^{(1)} + \beta^{(2)} \right)$,
with
\begin{eqnarray}
\label{eq:ctH1}
\delta_H^{(1)} & = & \frac{1}{i{(2\pi)}^d} \frac{g^2}{M_W^2}
                     I_1(m_H^2;1) \, \frac{\ep}{8\,(\ep-4)}
\\
\label{eq:ctH2}
\delta_H^{(2)} & = & {\left[
                     \frac{1}{i{(2\pi)}^d} \frac{g^2}{M_W^2}
                     \right]}^2
\left\{
I_1(m_H^2;1)^2
          \left( 
            \frac{1}{128}
          - \frac{3}{16\,(\ep-4)}
          + \frac{1}{8\,(\ep-4)^2}
          \right)
\right. \nonumber \\ &&
       {} + m_H^2 \, I_2(m_H^2,m_H^2,m_H^2;1,1,1)
          \left(
            \frac{3\,\ep}{64}
          + \frac{9}{64}
          + \frac{3}{8\,(\ep-4)}
          \right)
\nonumber \\ && \left.
       {} + m_H^2 \, I_2(m_H^2,0,0;1,1,1)
          \left(
            \frac{\ep}{64}
          + \frac{11}{64}
          + \frac{3}{4\,(\ep-4)}
          \right)
\right\}
\nonumber \\ &&
       {} + \frac{1}{i{(2\pi)}^d} \frac{g^2}{M_W^2}
          I_1(m_H^2;1)
          \left(
            \frac{1}{4}
          + \frac{\ep}{8}
          + \frac{1}{(\ep-4)}
          \right) \delta_{m_H}^{(1)}
\\
\label{eq:ctbt1}
\beta^{(1)} & = &  \frac{1}{i{(2\pi)}^d} \frac{g^2}{M_W^2}
\left\{
 - \frac{3}{8} I_1(m_H^2;1)
 - \frac{1}{4} I_1(M_W^2;1)
 - \frac{1}{8} I_1(M_Z^2;1)
\right\}
\\
\label{eq:ctbt2}
\beta^{(2)} & = & {\left[
                     \frac{1}{i{(2\pi)}^d} \frac{g^2}{M_W^2}
                     \right]}^2
\left\{
            I_1(m_H^2;1)^2
          \left(
          - \frac{3}{32}
          - \frac{3}{8\,(\ep-4)}
          \right)
\right. \nonumber \\ &&
       {} + m_H^2 \, I_2(m_H^2,m_H^2,m_H^2;1,1,1)
          \left(
          \frac{15}{64}
        - \frac{9\,\ep}{64}
          \right)
\nonumber \\ &&
       {} + I_2(m_H^2,0,0;1,1,1)
          \left(
          m_H^2 \frac{9-3\,\ep}{64}
        - M_W^2 \frac{(2\,c_W^2+1)(\ep-2)(\ep-9)}{32\,c_W^2}
          \right)
\nonumber \\ && \left.
       {} + I_1(m_H^2;1) I_1(M_W^2;1)
          \left(
            \frac{3\,\ep}{32}
          - \frac{3}{16}
          \right)
       {} + I_1(m_H^2;1) I_1(M_Z^2;1)
          \left(
            \frac{3\,\ep}{64}
          - \frac{3}{32}
          \right)
\right\}
\nonumber \\ &&
       {} + \frac{1}{i{(2\pi)}^d} \frac{g^2}{M_W^2}
\left\{
            I_1(m_H^2;1)
          \left(
            \frac{3}{2}
          - \frac{3\,\ep}{8}
          \right)
   {} + \frac{1}{2} I_1(M_W^2;1)
      + \frac{1}{4} I_1(M_Z^2;1)
\right\} \delta_{m_H}^{(1)}
\nonumber \\ &&
\end{eqnarray}
The one- and two-loop scalar integrals used in these expressions are defined
by
\begin{eqnarray}
\label{eq:I2}
\lefteqn{
    I_2(m_1^2,m_2^2,m_3^2\,; n_1,n_2,n_3) =
}\hspace{1cm}  \nn \\ &&
      \int {d^dk_1\, d^dk_2 }\,\,
           P{(k_{1}\,;\,m_{1})}^{n_1} 
           P{(k_2\,;\,m_{2})}^{n_2}
           P{(k_{1}+k_{2}\,;\,m_{3})}^{n_3}
\\
\label{eq:I1}
\lefteqn{
       I_1(m_1^2\,; n_1) = 
}\hspace{1cm}  \nn \\ &&
         \int {d^dk_1}\,\,
         P{(k_{1}\,;\,m_{1})}^{n_1}                 
\end{eqnarray}
with
\begin{equation}
 P{(k\,;\,m)} = \frac{1}{k^2+m^2}\,.
\end{equation}
Note that in $\beta^{(1)}$ and $\beta^{(2)}$, it is necessary
to keep not only the leading power of $m_H$, but also the
next-to-leading power.

The Higgs mass renormalization constant
$Z_{m_H}=1-\delta_{m_H}^{(1)}-\delta_{m_H}^{(2)}$~\cite{Borodulin:1996br}
\begin{eqnarray}
\label{eq:ctmh1}
\delta_{m_H}^{(1)} & = &
                   {\left(\frac{g^2}{16\pi^2}\right)}
                   {\left(\frac{m_H^2}{4\pi}\right)}^{-\frac{\ep}{2}}
                   \Gamma\left(1+\frac{\ep}{2}\right)
                   \frac{m_H^2}{M_W^2}
\left\{
          - \frac{3}{2\,\ep}
          - \frac{3}{2}
          + \frac{3}{16}\pi\sqrt{3}
\right. \nonumber \\ && \left.
    {}  + \ep \left(
          - \frac{3}{32}\pi\sqrt{3}\log{3}
          + \frac{3}{16}\pi\sqrt{3}
          + \frac{1}{16}\pi^2
          + \frac{3}{8}\sqrt{3}\,C
          - \frac{3}{2}
           \right)
\right\} \, ,
\\
\label{eq:ctmh2}
\delta_{m_H}^{(2)} & = &
                   {\left(\frac{g^2}{16\pi^2}\right)}^2
                   {\left(\frac{m_H^2}{4\pi}\right)}^{-\ep}
                   \Gamma^2\left(1+\frac{\ep}{2}\right)
                   \frac{m_H^4}{M_W^4}
\left\{
          - \frac{27}{8\,\ep^2}
          +\frac{1}{\ep} \left(
            \frac{27}{32}\pi\sqrt{3}
          - \frac{363}{64}
            \right)
\right. \nonumber \\ && \left.
          - \frac{1575}{256}
          - \frac{27}{64}\pi\sqrt{3}\log{3}
          + \frac{291}{128}\pi\sqrt{3}
          - \frac{39}{32}\pi\,C
          - \frac{177}{512}\pi^2
          + \frac{3}{32}\sqrt{3}\,C
          + \frac{63}{32}\zeta(3)
\right\} \, .
\nonumber \\
\end{eqnarray}




\begin{thebibliography}{99}

\bibitem{LEP:2004}
The LEP Collaborations ALEPH, DELPHI, L3, OPAL, the LEP Electroweak
   Working Group, the SLD Electroweak and Heavy Flavor Groups,
arXiv:hep-ex/0412015.

\bibitem{Veltman:1976rt}
M.~J.~G.~Veltman,
Acta Phys.\ Polon.\ B {\bf 8} (1977) 475.

\bibitem{vanderBij:1983bw}
J.~van der Bij and M.~J.~G.~Veltman,
Nucl.\ Phys.\ B {\bf 231} (1984) 205.

\bibitem{vanderBij2}
J.~J.~van der Bij,
Nucl.\ Phys.\ B {\bf 248} (1984) 141;\\
J.~J.~van der Bij,
Nucl.\ Phys.\ B {\bf 267} (1986) 557.

\bibitem{Barbieri:1993ra}
R.~Barbieri, P.~Ciafaloni and A.~Strumia,
Phys.\ Lett.\ B {\bf 317} (1993) 381.

\bibitem{akhoury}
R.~Akhoury, J.~J.~van der Bij and H.~Wang,
Eur.\ Phys.\ J.\ C {\bf 20} (2001) 497
[arXiv:hep-ph/0010187].

\bibitem{Boughezal:2004ef}
R.~Boughezal, J.~B.~Tausk and J.~J.~van der Bij,
arXiv:hep-ph/0410216.

\bibitem{Awramik:2004ge}
M.~Awramik, M.~Czakon, A.~Freitas and G.~Weiglein,
Phys.\ Rev.\ Lett.\  {\bf 93} (2004) 201805
[arXiv:hep-ph/0407317].

\bibitem{Awramik:2003rn}
M.~Awramik, M.~Czakon, A.~Freitas and G.~Weiglein,
Phys.\ Rev.\ D {\bf 69} (2004) 053006
[arXiv:hep-ph/0311148].

\bibitem{Peskin}
M.~E.~Peskin and T.~Takeuchi,
Phys.\ Rev.\ Lett.\  {\bf 65} (1990) 964;\\
M.~E.~Peskin and T.~Takeuchi,
Phys.\ Rev.\ D {\bf 46} (1992) 381.

\bibitem{Einhorn:1988tc}
M.~B.~Einhorn and J.~Wudka,
Phys.\ Rev.\ D {\bf 39} (1989) 2758.

\bibitem{qgraf}
P.~Nogueira,
J.\ Comput.\ Phys.\  {\bf 105} (1993) 279.

\bibitem{FORM}
J.~A.~M.~Vermaseren,
arXiv:math-ph/0010025.

\bibitem{smirnovbook}
V.~A.~Smirnov,
``Applied asymptotic expansions in momenta and masses,''
Springer-Verlag, Berlin (2002).

\bibitem{Anastasiou:2004vj}
C.~Anastasiou and A.~Lazopoulos,
JHEP {\bf 0407}, 046 (2004)
[arXiv:hep-ph/0404258].

\bibitem{Broadhurst:1998rz}
D.~J.~Broadhurst,
Eur.\ Phys.\ J.\ C {\bf 8} (1999) 311
[arXiv:hep-th/9803091].

\bibitem{Fleischer:1999mp}
J.~Fleischer and M.~Y.~Kalmykov,
Phys.\ Lett.\ B {\bf 470} (1999) 168
[arXiv:hep-ph/9910223].

\bibitem{Antonelli}
F.~Antonelli, M.~Consoli and G.~Corbo,
Phys.\ Lett.\ B {\bf 91} (1980) 90;\\
F.~Antonelli, G.~Corbo, M.~Consoli and O.~Pellegrino,
Nucl.\ Phys.\ B {\bf 183} (1981) 195.

\bibitem{Veltman:1980fk}
M.~J.~G.~Veltman,
Phys.\ Lett.\ B {\bf 91} (1980) 95.

\bibitem{Longhitano:1980iz}
A.~C.~Longhitano,
Phys.\ Rev.\ D {\bf 22} (1980) 1166.

\bibitem{Marciano:1990dp}
W.~J.~Marciano and J.~L.~Rosner,
Phys.\ Rev.\ Lett.\  {\bf 65} (1990) 2963
[Erratum-ibid.\  {\bf 68} (1992) 898].

\bibitem{Altarelli:1990zd}
G.~Altarelli and R.~Barbieri,
Phys.\ Lett.\ B {\bf 253} (1991) 161.

\bibitem{Borodulin:1996br}
V.~Borodulin and G.~Jikia,
Phys.\ Lett.\ B {\bf 391} (1997) 434
[arXiv:hep-ph/9609447].

\end{thebibliography}
\end{document}